\documentclass{article}
\usepackage{spconf,amsmath,graphicx,hyperref,enumitem,subcaption}
\usepackage{booktabs}
\usepackage{comment}
\usepackage{multirow}
\usepackage{xcolor}

\title{AT-Differ: Dataset for Audio--Text Difference Captioning}
\name{Akihiro Fujii$^1$, Yuki Okamoto$^1$, Yuki Saito$^{1,2}$, and Hiroshi Saruwatari$^1$}
\address{$^1$The University of Tokyo, Japan, \\ $^2$National Institute of Advanced Industrial Science and Technology (AIST), Japan}
\begin{document}
\ninept
\maketitle
\begin{abstract}
Evaluating audio--text semantic alignment is important in various audio--language tasks.
Existing methods mainly quantify audio--text semantic alignment using a single score and cannot explicitly describe how audio and text differ.
In this paper, we propose audio--text difference captioning, a task that generates textual descriptions of differences between audio and text.
We also construct a new dataset \textit{AT-DIFFER} containing audio--text pairs with systematically simulated differences and corresponding difference captions.
Using \textit{AT-DIFFER}, we fine-tune an existing multimodal large language model to describe differences between audio and text.
The experimental results show that the original model has limited ability to describe such differences, whereas fine-tuning on \textit{AT-DIFFER} substantially improves this ability.
\end{abstract}
%
\section{Introduction}
\label{sec:intro}

Research on text-to-audio generation (TTA), which aims to generate general audio from text descriptions, has attracted increasing attention~\cite{Liu_TASLP_2024,Majumder_ACMM_2024}. 
In TTA, it is crucial that the generated audio faithfully reflects the content of the input text.
Audio--text semantic alignment is also crucial in other audio--language tasks, such as audio captioning (AC)~\cite{Drossos_WASPAA_2017} and text-queried audio retrieval~\cite{oncescu_interspeech_2021,Munakata_ICASSP_2025}.
Furthermore, understanding audio--text semantic alignment is also important in rapidly advancing multimodal large language models (MLLMs)~\cite{Tang_ICLR_2024,Xu_arXiv_2025,KimiTeam_arXiv_2025}.

Several methods have been proposed to automatically predict audio--text semantic-alignment scores, showing reasonable correlations with human subjective evaluations~\cite{kanamori_interspeech_2025,Takano_APSIPA_2025}.
However, these methods only predict semantic-alignment scores and do not reveal what types of misalignment occur between audio and text.
A method using MLLMs has also been proposed to generate explanations for audio--text semantic-alignment scores~\cite{Liang_ICLR_2024}. 
However, whether the textual reasoning for score prediction accurately describes audio--text semantic misalignment has not been evaluated.

In this paper, we propose audio--text difference captioning, a task that generates captions describing the differences between audio and text.
Audio--text difference captioning provides more detailed and informative explanations of audio--text semantic alignment than a single score.
In TTA, we can obtain evidence-based explanations regarding the quality of the generated audio, which may make it possible to design models that utilize this as feedback.
In AC, this serves as an evaluation of the generated text, and in audio retrieval, it provides explanations that allow users to understand candidate audio samples without actually listening to them, thereby contributing to improved usability.

This paper makes the following contributions:
\textbf{(1)} We construct an open-source dataset\footnote{https://github.com/sarulab-speech/AT-DIFFER\label{footnote_github}} called \textit{AT-DIFFER} (\textbf{A}udio--\textbf{T}ext \textbf{DIFFER}ence captioning) for describing differences between audio and text;
\textbf{(2)} We fine-tune an existing MLLM, Qwen3-Omni, using AT-DIFFER to generate audio--text difference captions and conduct a benchmark analysis.

\begin{figure}[t]
  \centering
  \includegraphics[width=\columnwidth]{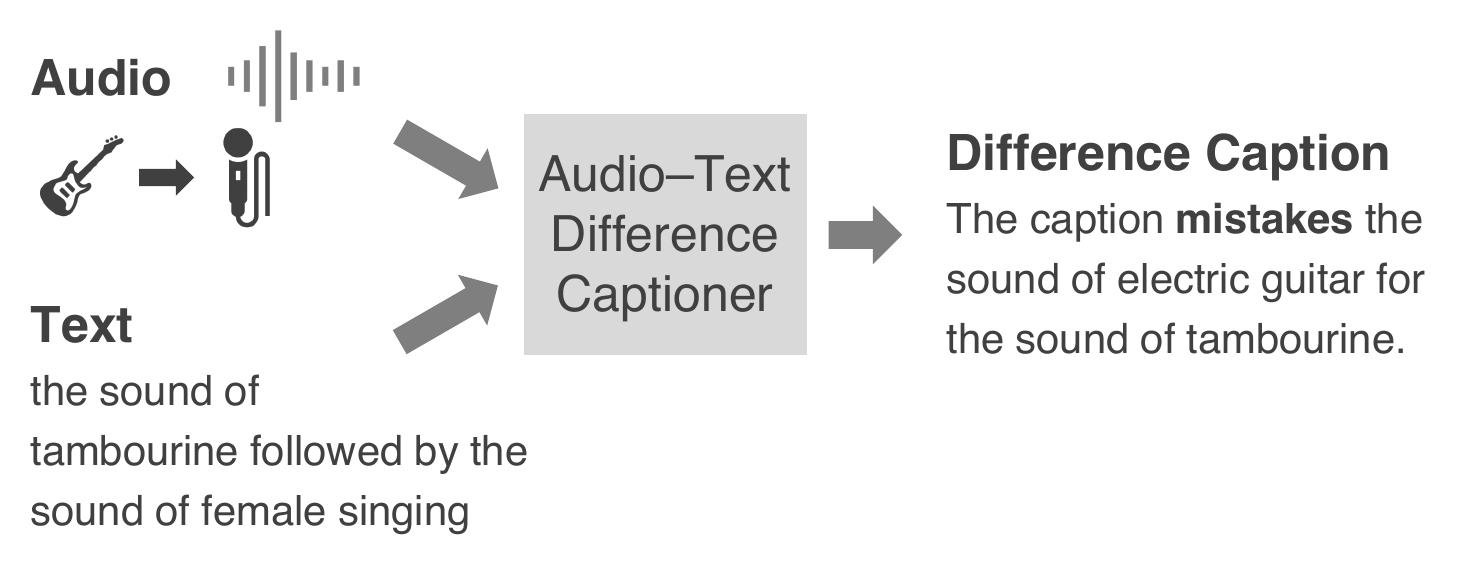}
  \caption{Overview of audio--text difference captioning.}
  \label{fig:task}
\end{figure}
\begin{table*}[t]
    \centering
    \setlength{\tabcolsep}{4pt}
    \caption{Templates of audio--text difference caption. ``Sim.'' and  ``Seq.'' denote ``Simultaneous'' and ``Sequential,'' respectively.}
    \label{tab:caption_template}
    \begin{tabular}{@{}ll|l@{}}
        \hline
        Audio structure & Difference type & Difference caption \\
        \hline
        \multirow{3}{6em}{Single, Sim., \& Seq. (common)} & Semantic difference & The caption mistakes the sound of [\textit{correct event}] for the sound of [\textit{wrong event}]. \\
         & Text omission & The caption leaves out the sound of [\textit{absent event}] that is actually present. \\
         & Sound omission & The caption adds the sound of [\textit{nonexistent event}] that is not actually there. \\
        \hline
        \multirow{2}{*}{Seq. only} & Seq. discrepancy & The caption reverses the order of  [\textit{event A}] and [\textit{event B}]. \\
         & Sim. discrepancy & The caption describes the sequential [\textit{event A}] and [\textit{event B}] as overlapping. \\
        \hline
        Sim. only & Sim. discrepancy & The caption describes the simultaneous [\textit{event A}] and [\textit{event B}] as sequential. \\
        \hline
    \end{tabular}%
\end{table*}
\section{Related Work}
\label{sec:related}

\textbf{Audio--text alignment evaluation.}
Several methods have been developed to evaluate audio--text semantic alignment.
Contrastive language-audio pretraining (CLAP)~\cite{ICASSP2023_Yusong_01} measures semantic alignment based on the similarity between audio and text embeddings.
Other methods predict audio--text semantic-alignment scores that correlate well with subjective human evaluations~\cite{kanamori_interspeech_2025,Takano_APSIPA_2025}.
Furthermore, a method has been proposed to classify the relationship between audio and text as entailment, neutral, or contradiction~\cite{Deshmukh_AAAI_2025}.
However, these approaches focus on quantifying or classifying audio--text semantic alignment rather than describing specific differences between audio and text.
In contrast, this paper aims to generate textual descriptions of such differences.

\textbf{Audio--text alignment types.}
A study has also defined different types of audio--text semantic alignment and applied them to the evaluation and training of TTA models~\cite{He_EMNLP_2025}.
These semantic-alignment types include temporal order, spatial distance, count, and compositionality, which also provide a useful basis for constructing our dataset.

\textbf{Cross-modal difference captioning.}
Several studies have explored the generation of textual descriptions of differences between modalities~\cite{Gordon_ECCV_2024,Hua_ECCV_2024}.
In the image domain, such approaches describe mismatches between generated images and their input text for evaluating text-to-image generation~\cite{Gordon_ECCV_2024}.
Our study extends this concept to the audio domain through audio--text difference captioning.

\textbf{MLLM-based alignment evaluation.}
A method using MLLM has also been proposed to evaluate differences between modalities.
Liang et al.\ used MLLMs to predict semantic-alignment scores across modalities and generated descriptions of cross-modal inconsistencies through a chain-of-thought reasoning process~\cite{Liang_ICLR_2024}.
Although this approach of generating textual explanations for the predictions is conceptually similar to ours, the generated explanations were not evaluated.
Therefore, it remains unclear how accurately MLLMs can identify and describe the differences between audio and text.
\begin{figure}[t]
    \centering    \includegraphics[width=0.95\columnwidth]{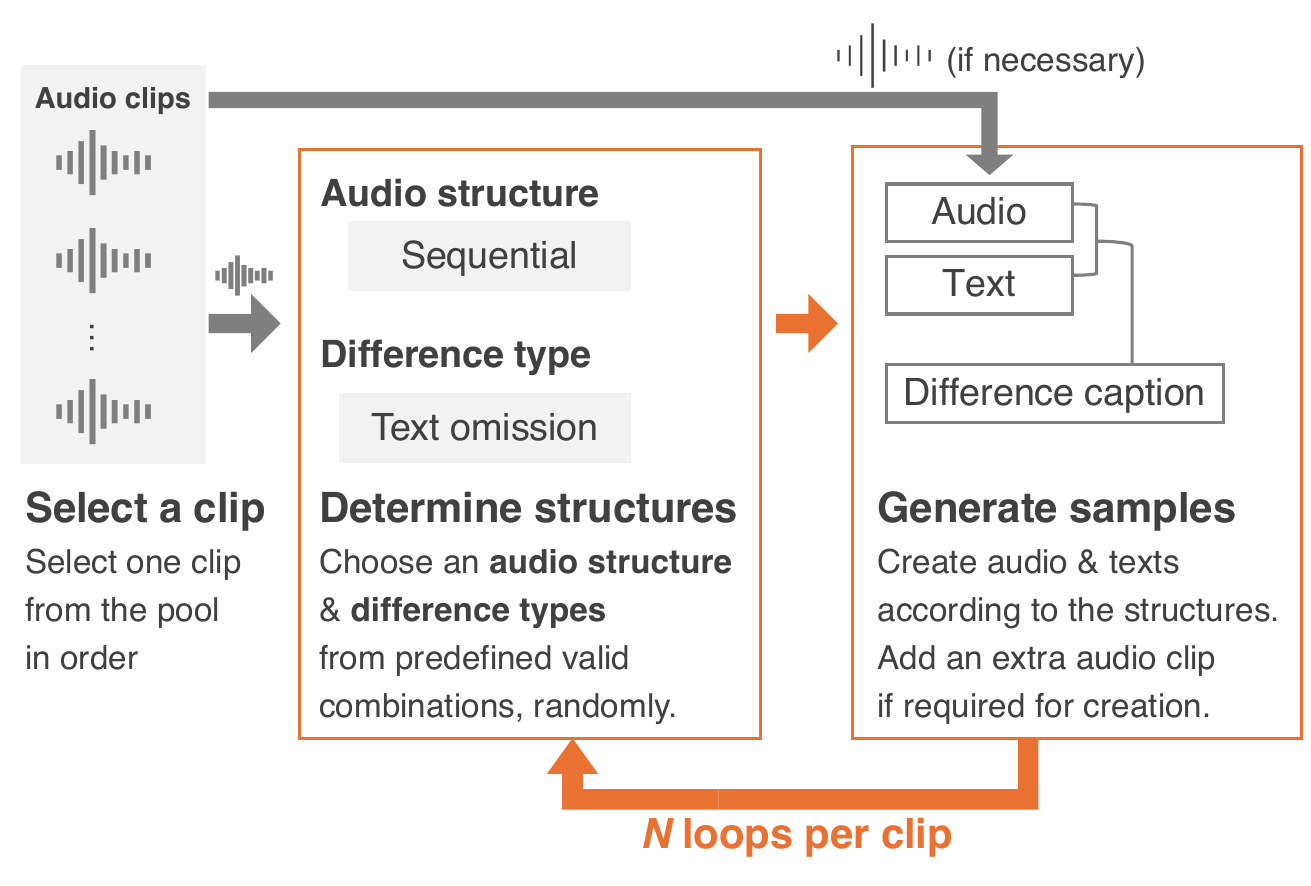}
    \caption{Procedure for constructing AT-DIFFER.}
    \label{fig:dataset_pipeline}
\end{figure}
\section{Creation of AT-DIFFER}
\label{sec:dataset}
\subsection{Dataset overview}
\label{subsec:dataset_overview}
Our dataset consists of the following contents.
\begin{itemize}
    \item \textbf{Audio--text pairs.} Audio–text pairs with various types of simulated semantic misalignment.
    All text is written in English, and the audio clips are provided as monaural, 32-bit float, 22.05~kHz RIFF WAV format.
    \item \textbf{Difference captions.} Texts describing the differences between the audio and text in each pair.
\end{itemize}

\subsection{Creation process}
Figure \ref{fig:dataset_pipeline} illustrates the dataset creation process. 
We use FSD50K-Solo~\cite{Yang_EUSIPCO_2026}, a subset of FSD50K~\cite{Fonseca_TASLP_2021} consisting of clips containing a single sound event, as the source audio dataset. 
For each clip, we randomly select $N = 6$ conditions from predefined valid combinations of audio structures and difference types.
For each condition, an additional sound event is randomly selected if necessary, and the audio--text pair and difference caption are generated accordingly. 
Texts and difference captions are generated from predefined templates by substituting sound event names with the corresponding labels provided by FSD50K.
Table~\ref{tab:caption_template} shows templates of audio--text difference captions.

\textbf{Audio structure.}
We consider three audio structures based on the composition and temporal relationships of sound events:
\begin{itemize}
\item \textbf{Single.} A single sound event in an audio clip.
\item \textbf{Simultaneous.} Two audio clips are superimposed so that the two sound events occur simultaneously.
\item \textbf{Sequential.} Two audio clips are concatenated so that the two sound events occur sequentially.
\end{itemize}
Thus, each audio clip contains one or two sound events. 

\textbf{Difference type.}
We consider five types of differences between the audio and text, categorized as content and temporal differences:
\begin{itemize}
\item \textbf{Semantic difference (Content).} A sound event in the audio is described as a different sound event.
\item \textbf{Text omission (Content).} A sound event present in the audio is omitted from the text.
\item \textbf{Sound omission (Content).} A sound event described in the text is absent from the audio.
\item \textbf{Sequential discrepancy (Temporal).} The order of two sequential sound events is reversed in the text.
\item \textbf{Simultaneity discrepancy (Temporal).} The temporal relationship between two events differs between the audio and text.
\end{itemize}
If the audio and text are semantically consistent with each other, the pair is labeled as ``no difference.’’

As an initial investigation, we limit each audio sample to at most two sound events and each audio--text pair to at most two difference types to reduce task complexity.
Our dataset construction process can be extended to three or more sound events and difference types.
Differences related to duration, count, and spatial information are excluded from this dataset.

\subsection{Splitting and statistics}
The training, validation, and test sets follow the original FSD50K splits, and the dataset creation procedure is applied independently to each split.
This yields $149{,}172$ training pairs ($30{,}770$ single, $51{,}279$ simultaneous, $67{,}123$ sequential), $14{,}466$ validation pairs ($3{,}034$ single, $4{,}941$ simultaneous, $6{,}491$ sequential), and $33{,}642$ test pairs ($6{,}982$ single, $11{,}604$ simultaneous, $15{,}056$ sequential).

\begin{figure}[t]
  \centering
  \includegraphics[width=\columnwidth]{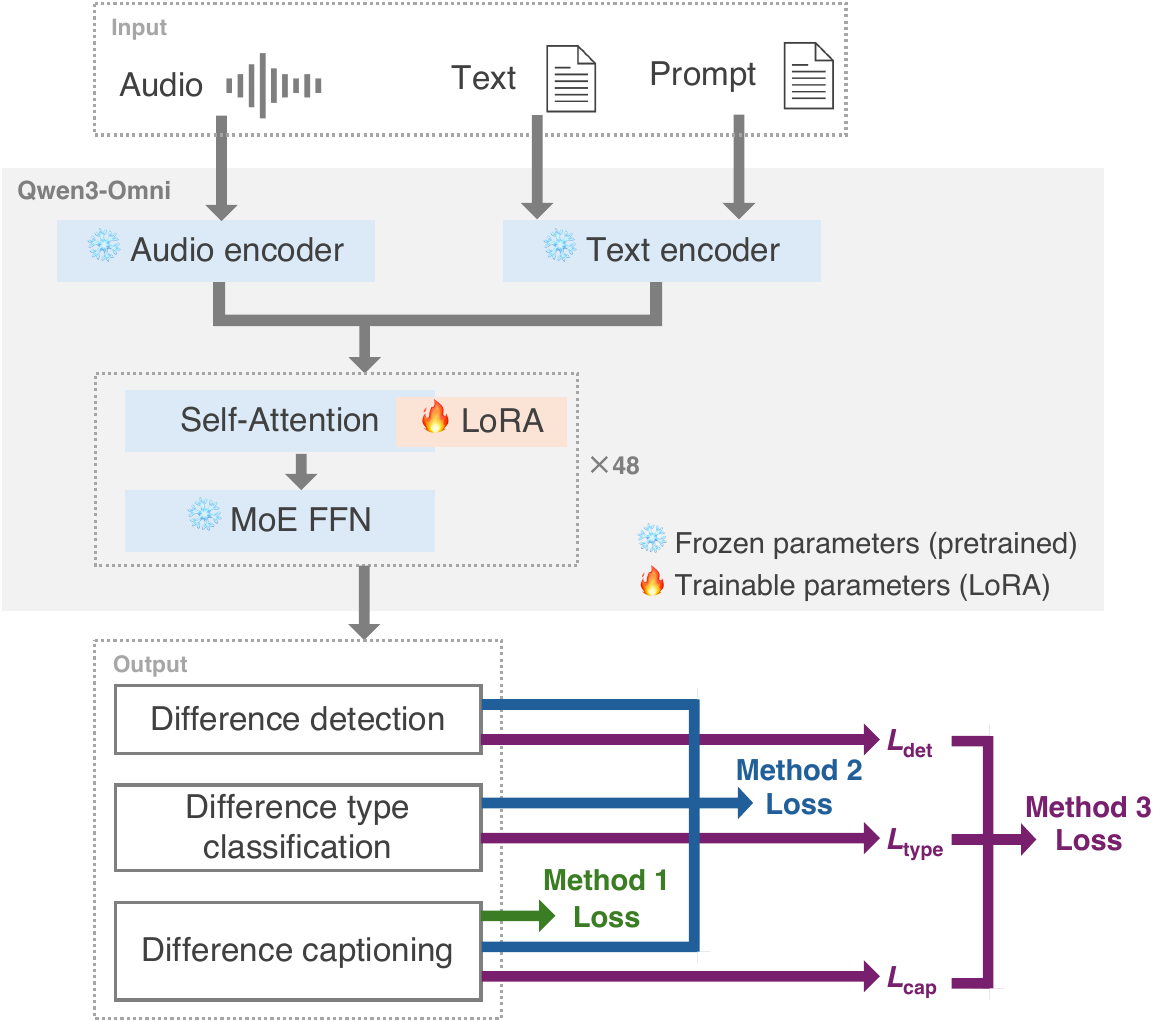}
  \caption{Model architecture and fine-tuning strategies.
  Parameters of self-attention projection matrices are fine-tuned using LoRA. Other parameters, such as those of encoders and Mixture-of-Experts feed-forward network (MoE FFN), are frozen.}
  \label{fig:architecture}
\end{figure}
%
\section{Benchmarking Audio--Text Difference Captioning}
\label{sec:benchmark}
\subsection{Model architecture and fine-tuning strategies}
\label{subsec:model_architecture}
Using AT-DIFFER constructed in Section \ref{sec:dataset},
we fine-tuned an existing MLLM, Qwen3-Omni-30B-A3B-Thinking\footnote{https://huggingface.co/Qwen/Qwen3-Omni-30B-A3B-Thinking}, and evaluated its performance on the following three tasks:
(i) \textbf{Difference detection}, binary classification of whether differences exist between the audio and text;
(ii) \textbf{Difference type classification}, classification of the differences between the audio and text into predefined types; and
(iii) \textbf{Audio--text difference captioning}, generation of textual descriptions of the differences between the audio and text.
The model is prompted to solve these three tasks sequentially to improve caption quality. 
This design builds upon Gordon et al. \cite{Gordon_ECCV_2024}, where models detect and describe differences in text--image pairs, and joint multi-task training was shown to outperform single-task training.

\begin{figure}
    \centering
    \includegraphics[width=0.9\columnwidth]{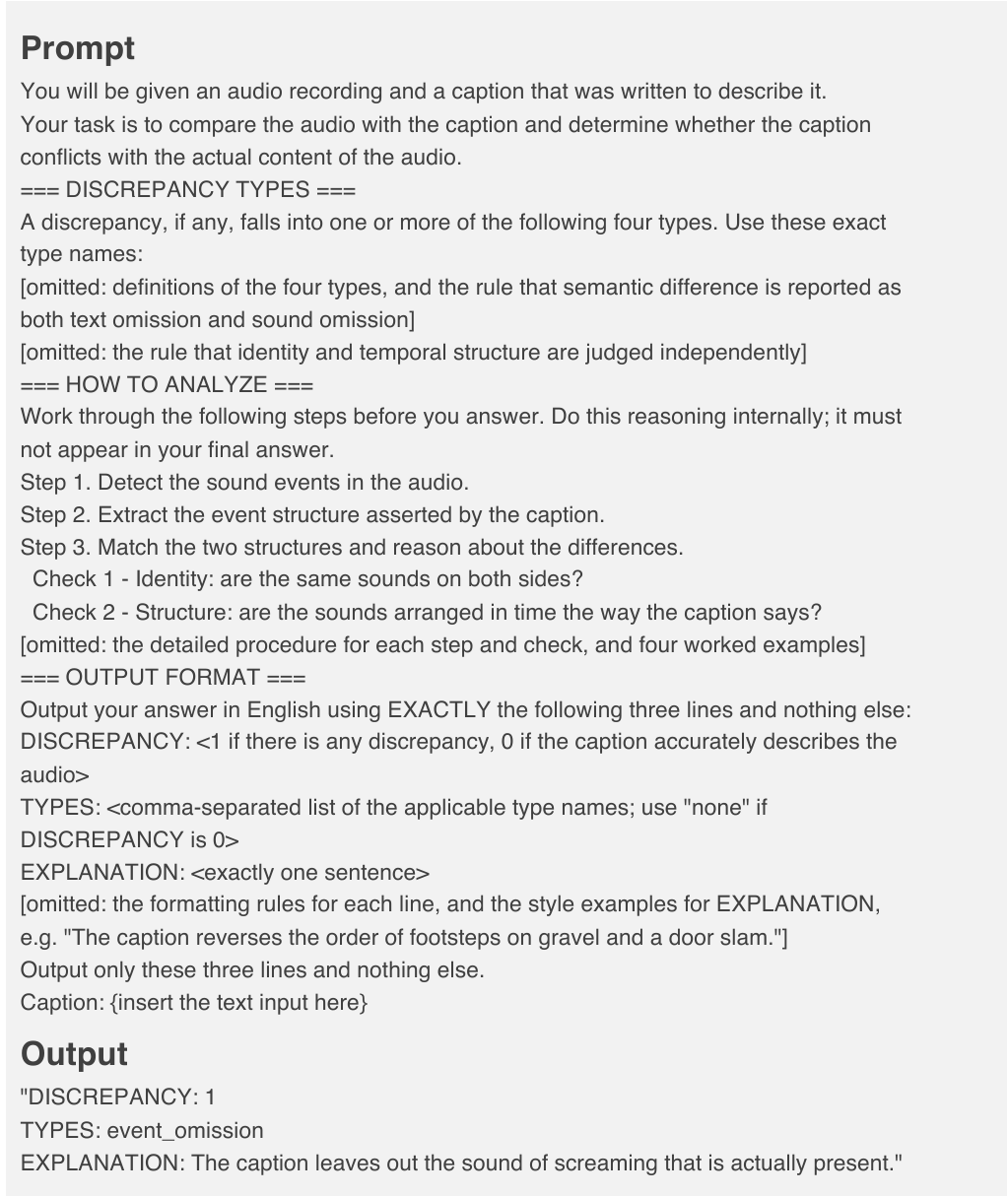}
    \caption{The abridged prompt and an example of model output.}
    \label{fig:prompt_and_output}
\end{figure}

Figure \ref{fig:architecture} shows the model architecture and fine-tuning strategies.
For each audio--text pair, we append the text to the prompt shown in Fig.~\ref{fig:prompt_and_output} and feed it to the MLLM together with the audio.
The model then generates outputs for tasks (i)–(iii) in the format specified in the prompt.
Fine-tuning is performed with three methods, using Low-Rank Adaptation (LoRA) applied to self-attention; all other weights are frozen.
The three methods differ in which outputs of tasks (i)–(iii) are supervised and how their losses are combined.
In all methods, the token-level loss is averaged within each audio–text pair and then across pairs, ensuring that each pair contributes equally regardless of output length.

\textbf{Method 1 (Difference caption only)}.
The loss is computed only between the model output for task (iii) and the ground-truth difference caption.

\textbf{Method 2 (Concatenate three tasks)}.
The outputs for tasks (i)–(iii) are concatenated, and a single cross-entropy loss is computed over all tokens.
Thus, each task’s contribution to the loss is proportional to its number of output tokens.

\textbf{Method 3 (Separate three tasks)}.
The loss is computed separately for each task, and the resulting losses are combined using a weighted sum. 
Let $\mathcal{L}_{\mathrm{det}}$, $\mathcal{L}_{\mathrm{type}}$, and $\mathcal{L}_{\mathrm{cap}}$ denote the cross-entropy losses for difference detection, difference type classification, and difference captioning, respectively. 
The final loss function is their weighted sum:
$\mathcal{L} = \lambda_{\mathrm{det}}\mathcal{L}_{\mathrm{det}} + \lambda_{\mathrm{type}}\mathcal{L}_{\mathrm{type}} + \lambda_{\mathrm{cap}}\mathcal{L}_{\mathrm{cap}}$.
In contrast to method 2, the contribution of each task to the loss can be adjusted through the weights $\lambda_{*}$.
\begin{table*}[t]
    \centering
    \setlength{\tabcolsep}{4pt}
    \caption{Results of audio--text difference captioning. 
    Bold indicates the best performance for each metric among the models. 
    }
    \label{tab:caption_results}
    \resizebox{\textwidth}{!} {%
    \begin{tabular}{@{}l cccc c c c c c ccc c@{}}
        \toprule
        \multirow{2}{*}{Model} & \multicolumn{4}{c}{BLEU\,$\uparrow$} & \multirow{2}{*}{METEOR\,$\uparrow$} & \multirow{2}{*}{ROUGE-L\,$\uparrow$} & \multirow{2}{*}{CIDEr-D\,$\uparrow$} & \multirow{2}{*}{SPICE\,$\uparrow$} & \multirow{2}{*}{SPIDEr\,$\uparrow$} & \multicolumn{3}{c}{BERTScore\,$\uparrow$} & \multirow{2}{*}{SBERT\,$\uparrow$} \\
        \cmidrule(lr){2-5} \cmidrule(lr){11-13}
         & 1 & 2 & 3 & 4 & & & & & & P & R & F1 &  \\
        \midrule
        Baseline & 0.550 & 0.429 & 0.337 & 0.269 & 0.273 & 0.498 & 1.021 & 0.449 & 0.735 & 0.545 & 0.539 & 0.542 & 0.735 \\
        Random \\
        ~~~Fixed audio structure \& diff. type & 0.768 & 0.687 & 0.614 & 0.539 & 0.370 & 0.779 & 0.892 & 0.377 & 0.634 & 0.676 & 0.672 & 0.675 & 0.627 \\
        ~~~Fixed sound event & 0.675 & 0.544 & 0.444 & 0.376 & 0.341 & 0.652 & 4.839 & 0.480 & 2.659 & 0.631 & 0.612 & 0.622 & 0.864 \\
        \midrule
        \textbf{Method 1} (Difference caption only) & 0.618 & 0.558 & 0.506 & 0.469 & 0.339 & 0.618 & 2.858 & 0.469 & 1.664 & 0.643 & 0.600 & 0.620 & 0.755 \\
        \textbf{Method 2} (Concatenate three tasks) & \textbf{0.915} & \textbf{0.892} & \textbf{0.867} & \textbf{0.843} & \textbf{0.573} & 0.900 & \textbf{6.882} & 0.809 & \textbf{3.845} & 0.899 & \textbf{0.889} & 0.894 & \textbf{0.911} \\
        \textbf{Method 3} (Separate three tasks) & 0.913 & 0.889 & 0.865 & 0.841 & 0.572 & \textbf{0.902} & 6.873 & \textbf{0.810} & 3.842 & \textbf{0.902} & \textbf{0.889} & \textbf{0.895} & \textbf{0.911} \\
        \bottomrule
    \end{tabular}%
    }
\end{table*}

\subsection{Experimental setup}
\label{subsubsec:experimental_setup}
\textbf{Model configuration.}
The audio clips were resampled from 22.05~kHz to 16~kHz.
The fine-tuning settings were common to all three methods.
LoRA was applied only to the query, key, value, and output projection matrices in the self-attention layers, with a rank of $16$ and a scaling factor of $\alpha = 32$.
Fine-tuning was conducted for 1 epoch, with an effective batch size of $32$.
The learning rate was set to $1.0 \times 10^{-4}$ and
decay was implemented using a cosine scheduler with a warmup ratio of $0.03$.
In both training and inference, an empty thinking block was prefilled at the beginning of the Qwen3-Omni-30B-A3B-Thinking response to shorten the output and avoid parsing failures caused by unterminated reasoning.
In method 3, in which the loss function consists of three losses, all weights $\lambda_{*}$ were set to $1/3$ in this experiment.
All experiments were run on a single node with two NVIDIA RTX PRO 6000 Blackwell Max-Q Workstation Edition GPUs (96 GB each).

\textbf{Evaluation metrics.}
The fine-tuned models are compared with the pre-trained baseline using the following metrics.
The performance of both difference detection and difference type classification is evaluated using accuracy.
For audio--text difference captioning, we use evaluation metrics commonly used in audio captioning~\cite{Mei_EURASIP_2022}, 
including BLEU~\cite{Papineni_ACL_2002}, METEOR~\cite{Banerjee_ACL_2005}, ROUGE-L~\cite{Lin_2004}, CIDEr~\cite{vedantam_CVPR_2015}, SPICE~\cite{Anderson_ECCV_2016}, SPIDEr~\cite{Liu_ICCV_2017}, BERTScore (precision, recall, F1-score)~\cite{Zhang_arXiv_2019}, and Sentence-BERT (SBERT)~\cite{Reimers_EMNLP_2019}.
Each metric is computed between the generated difference caption and the ground-truth difference caption.

\textbf{Compared methods.}
We compared the three fine-tuning strategies described in Sec.~\ref{subsec:model_architecture} with the baseline model and two types of randomly simulated difference captions as follows:
\begin{itemize}
\item \textbf{Baseline.}
We used the original Qwen3-Omni-30B-A3B-Thinking model without fine-tuning.
\item \textbf{Random (Fixed audio structure \& diff. type).}
We simulated difference captions by modifying the ground-truth difference captions, preserving the audio structure and difference type while replacing each sound event with a randomly selected one.
This provides chance-level scores for the metrics given correct identification of only the audio structure and the difference type.
\item \textbf{Random (Fixed sound event).}
We simulated difference captions by modifying the ground-truth difference captions, preserving each sound event while replacing the difference type with a randomly selected one.
This provides chance-level scores for the metrics given correct identification of only the sound events.
\end{itemize}

\subsection{Results and discussion}
\textbf{Difference detection.}
The overall accuracy was $80.9 \%$ for the baseline model, $75.4 \%$ for method 1, $94.8 \%$ for method 2, and $94.7 \%$ for method 3.
Although method 1 underperformed the baseline model, methods 2 and 3 outperformed it, achieving the highest overall accuracy.
In method 1, the loss does not depend on the difference detection; this likely leads to the degraded detection performance.

\begin{figure}[t]
    \centering
    \includegraphics[width=\columnwidth]{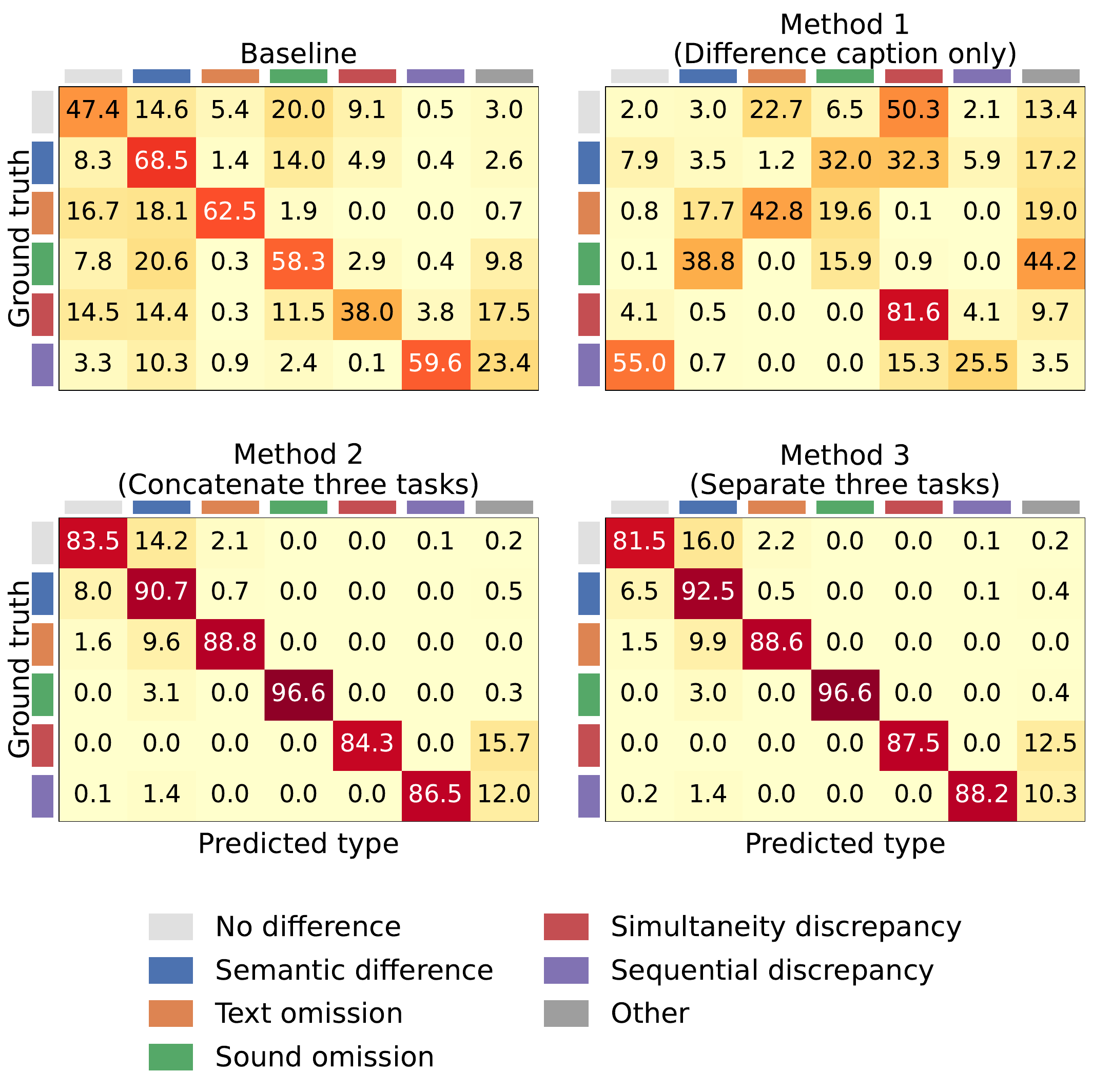}
    \caption{Confusion matrices for difference type classification. 
    The matrices cover only samples annotated with at most one difference type;
    predictions assigning two or more types are counted as ``other.''}
    \label{fig:confusion}
\end{figure}

\textbf{Difference type classification.}
The baseline model and methods 1, 2, and 3 achieved overall accuracies of $58.5 \%$, $20.2 \%$, $91.7 \%$, and $91.9 \%$, respectively.
Figure~\ref{fig:confusion} shows the confusion matrix for difference type classification for samples with one or fewer differences.
From these results, methods 2 and 3 achieved higher classification accuracy than the baseline, whereas method 1 showed lower accuracy.

This result suggests that training focused solely on difference captioning may reduce the ability to classify difference types as well.
In contrast, in method 2 and method 3, which learn the difference detection, type, and caption together, the classification accuracy also improved.
The results demonstrate the effectiveness of learning three tasks jointly.

\textbf{Difference captioning.}
Table~\ref{tab:caption_results} shows the results.
All proposed methods outperformed the baseline model across all evaluation metrics.
Methods 2 and 3, which incorporate intermediate classifications into the training, consistently outperformed method 1, which trains only on the final output.
Moreover, methods 2 and 3 achieved higher scores than the randomly generated difference captions across all metrics.
This result suggests that learning both difference detection and difference type classification is effective for generating difference captions, and captions generated here are not just following the pattern of training data.

Regarding the difference between methods 2 and 3,
method 3 treats the three tasks as separate loss terms, whereas method 2 concatenates them into a single sequence.
Method 3 achieved performance comparable to method 2, showing no clear advantage of using separate loss terms. 
Overall, these results suggest that jointly learning difference detection, difference type classification, and difference captioning is effective, while separating their loss terms does not clearly improve performance.
\section{Conclusion}
\label{sec:conclusion}
In this paper, we proposed audio--text difference captioning, a task that generates textual descriptions of differences between audio and text, and constructed \textit{AT-DIFFER}, a new dataset with systematically simulated differences. 
We found that an existing MLLM has limited ability to describe differences between audio and text, whereas fine-tuning on \textit{AT-DIFFER} substantially improves this ability.
Furthermore, jointly learning difference detection, difference type classification, and difference captioning was more effective than training only on difference captioning.
Future work includes extending the dataset to more diverse differences and evaluating other MLLMs.     

\vfill\pagebreak

\section{Acknowledgment and Use of Generative AI Disclosure}
\vspace{-1pt}
This work was supported by JSPS KAKENHI Grant Number 25K21221, 26H02531, Support Center for Advanced Telecommunications Technology Research Foundation, JST Moonshot Grant Number JPMJMS2011, and JST BOOST JPMJBY24C9.

\textbf{Use of Generative AI Disclosure.} Anthropic’s Claude and Claude Code were used to assist with manuscript drafting and editing.

\bibliographystyle{IEEEbib}
\bibliography{strings,refs_original}

\end{document}